\font\tenit=cmti10
 1
 1
 1

\documentstyle[epsf,aps]{revtex}
\def\ee{\end{equation}}
\def\be{\begin{equation}}
\title{ Non-Markovian Persistence at the  PC point of a 1d non-equilibrium 
kinetic Ising
model} 
\author{\sl N\'ora Menyh\'{a}rd\\ {\tenit  Research Institute for Solid State
Physics,
 H-1525 Budapest,P.O.Box 49, Hungary}}
\author{\sl G\'eza \'Odor\\ {\tenit Research Institute for Materials Science,
H-1525 Budapest, P.O.Box 49, Hungary}}
\begin{document}
\maketitle
\vskip .3cm

\begin{abstract}
One-dimensional  non-equilibrium kinetic Ising models evolving under the 
competing effect
of spin flips at zero temperature and nearest neighbour spin exchanges
exhibiting a parity-conserving (PC) phase transition on the level of
kinks are investigated here  numerically from the point of view 
of the underlying spin system.
The dynamical persistency exponent $\Theta$ and  
the exponent $\lambda$ 
characterising the two-time autocorrelation function of the
total magnetization  under nonequilibrium conditions are reported.
It is found that the PC transition has strong effect:
the process becomes non-Markovian and the above exponents exhibit
 drastic changes as compared to the Glauber-Ising case.
\end{abstract}
\bigskip
\bigskip

In recent years , two nonequilibrium dynamical critical exponents
have been discovered, which arise under non-equilibrium conditions.
The non-equilibrium (short-time) exponent $\lambda$
 characterizes two-time correlations in systems relaxing to
their critical state in the process of quenching from infinitely
high temperatures to $T_c$ \cite{Jan89,Hus89}. Recently, one more
critical exponent was proposed \cite{god}, the persistence exponent 
$\Theta$,
associated with the probability $p(t)\propto t^{-\Theta}$, that the
global order parameter has not changed sign up to time $t$ after a
quench to the critical point \cite{MBCS96}. For some known examples, cited in
ref. \cite{MBCS96},
the scaling law 
\begin{equation}
\Theta Z={\lambda -d +1-{\eta\over 2}}
\end{equation}
 is satisfied
(here d is the dimensionality and $\eta$ is the static critical
exponent of the order parameter correlation function),
which has been derived assuming that the dynamics of the order
parameter is a Markovian process. In general, however, $\lambda$ and
$\Theta$ have been proposed\cite{MBCS96} to be independent, new critical 
dynamical
exponents.

One of the soluble examples is the $d=1$ Ising model with Glauber kinetics.
In this case the critical temperature is at $T=0$, and as shown in
ref.\cite{MBCS96}, the persistence exponent is $\Theta=1/4$
for the global order parameter which is the total magnetisation
$M(t)$. Moreover, $\lambda$ is known to be $\lambda=1$ in this model.
The aim of the present note is to study these new dynamical
critical exponents in a simple {\it non-equilibrium} Ising system
(NEKIM) introduced in
\cite{MN94}.The phase diagram of NEKIM consists -essentially -
of a line of (first order) Ising-type transitions, which line ends
at a specific point. At this endpoint, on the level of kinks
(phase boundaries) a second order phase transition takes place from an
absorbing to an active state which, however, belongs to the parity
conserving (PC) universality class \cite{gra84,gra89,MN94,jen94,dani,kim94}.
 The critical
fluctuations of this PC transition exert a pronounced effect on the 
underlying spin system as found earlier \cite{MEOD96} thus e.g.
the the classical dynamical exponent $Z$, defined, as usual through
$\tau\propto\xi^Z$ with $\xi\propto p^{-\nu}$ was found to be 
$Z=1.75(1)$ instead
of the Glauber-Ising value of $Z=2$. (We note here that 
$p=e^{-\frac{4J}{kT}}$ plays the role of $\frac{T-T_c}{T_c}$ in 1d 
with $T_c=0$ and the
static exponents are defined as powers of $p$ for $T\rightarrow 0$ ).

The question arises how the citical fluctuations of the PC transition
affect the other two
critical dynamical exponents $\Theta$ and $\lambda$.
Before entering into the details of our results for $\Theta$ and  
$\lambda$, the model will be described in some detail.
\vglue 0.5cm

In NEKIM the system evolves
under  a combined effect of spin-flips  and spin-exchanges.
The
spin-flip transition rate 
in one-dimension
for spin $s_i$ ($s_i=\pm1$)
sitting at site $i$  is \cite{gla63}:
\be
w_i = {\frac{\Gamma}{2}}(1+\delta s_{i-1}s_{i+1})\left(1 - 
{\gamma\over2}s_i(s_{i-1} + s_{i+1})\right) 
\ee
where $\gamma=\tanh{{2J}/{kT}}$ ($J$ denoting the coupling constant in
the Ising Hamiltonian), $\Gamma$ and $\delta$ are further
 parameters.
At  $T=0$ $\gamma=1$ and there are two independent non-zero rates
${\Gamma\over2}(1-\delta)$ and
 ${\Gamma\over2}(1+\delta)$, responsible for
 random walk, and pairwise annihilation 
 of kinks, respectively.\\
The spin-exchange transition rate of nearest neighbour spins
(the Kawasaki\cite{kaw72} rate at $T=\infty$) is \newline
$w_{ii+1}={1\over2}p_{ex}[1-s_is_{i+1}],$
where $p_{ex}$ is the probability of spin exchange.
 Spin-flip and spin-exchange have been applied
alternatingly.

In this system, at $T=0$, a PC   type
phase transition takes place. In \cite{MN94} we have started from a random
initial state and determined the phase boundary in the ($\delta, p_{ex}$) 
plane.
In the following we will choose a typical point on this phase diagram
and make simulations at  this point.
 The parameters chosen are:
$\Gamma=.35, p_{ex}=.3, \delta_c=-.395(2)$. 
 In the simulations the spin-flip part has been
applied using two-sublattice updating. After that
 we have stored the states of the spins  and made L
(L is the size of the system) random attempts of exchange using 
always the 
stored situation for the states of the spins before updating.
All these together has been counted as one time-step
of  updating. ( Usual MC update in this last step
enhances the effect of $p_{ex}$ and leads to $\delta_c=-.362(1)$).

In \cite{MBCS96} it has been argued that for studying non-equilibrium critical
dynamics, the global, rather than the local order parameter should be 
considered.  The non-equilibrium
nature of the problem under consideration  is partly due to the 
model itself and partly due to the
 conditions of a quench from $T=\infty$ to $T=0$.
 This latter means that
we will restrict ourselves to completely random initial states  
and follow the behaviour of 
the system using the rules described above.
The persistency exponent $\Theta$ is defined via the probability $p(t)$
that the {\it global} order parameter, which in our case
is the total magnetization: $<M_{k=0}(t)>=\frac{1}{L}<\sum_{i}s_i(t)>$,
  has not changed sign up to time
$t$:
\begin{equation}
p(t)\propto t^{-\Theta}
\label{1.1}
\end{equation}
Finite size scaling (FSS) applied to the persistence problem 
\cite{MBCS96} leads to the form:
\begin{equation}
p(t)=L^{-\Theta Z} g(t/{L^Z})
\label{1.2}
\end{equation}
\begin{figure}[ht]
  \centerline{\epsfxsize=10cm
                   \epsfbox{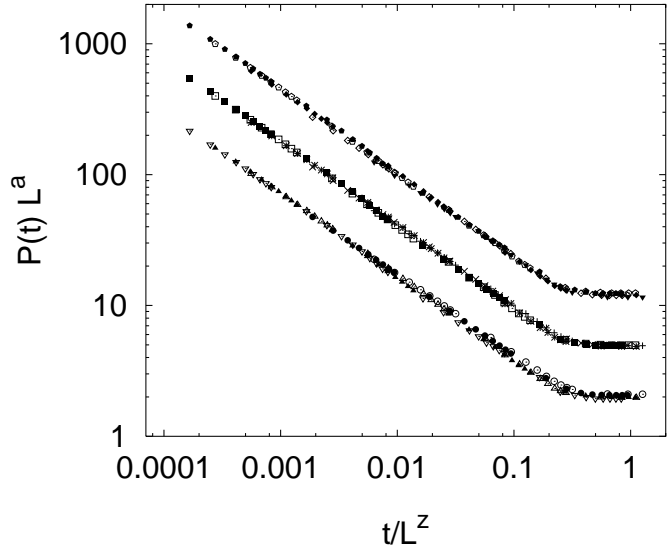}
                    \vspace*{6mm}
             }
   \caption{$p(t)L^a$, $a={\Theta Z}$, plotted against $t/L^Z$ 
 with $Z=1.75$ and $\Theta=.65,.67,.69$. For clarity the
$.65$ and $.69$ data have been multiplied and divided by a factor
of  $2$, respectively. The simulational data  exhibited here have been 
obtained for $L=100, 200, 400, 800$ sized systems labeled by
different symbols, for statistical averages between ${3\times 10^6}-
10^5$ samples.}
\end{figure}

Simulations have been carried out in the range $50\leq L\leq 2000$ 
with periodic as well as antiperiodic boundary conditions and at least for
$10^5$ independent runs.
Fig.1 shows those of our results which have the best statistics with averages
over up to $10^6$ independent random initial configurations
( those configurations, however, for which 
$M(0)=0$ exactly, were discarded) with periodic
boundary conditions 
 for $L=100,200,400$ and $800$. Using $Z=1.75(1)$ \cite{MEOD96} the best fit
corresponds to $\Theta=.67(1)$. The fact that periodic boundary conditions
allow only an even number
of kinks leads eventually to perfect ordering of spins  seen as levelling off
of curves on Fig.1. Our simulations  with  antiperiodic boundary
conditions have led to the same value of $\Theta$ as above though the form
of the scaling function $g(t/{L^Z})$ in this case is different, of course.
For comparison we have also simulated the exactly soluble Glauber-Ising
case and found the expected value of $\Theta=.25$ within the accuracy of
the simulations.

The local autocorrelation function defines the new exponent
$\lambda$ \cite{Jan89,Hus89}:
\begin{equation}
A(t,0)=\frac{1}{L}<\sum_{i}s_i(0)s_i(t)>\propto t^{-\frac{\lambda}{Z}}.
\end{equation}
We have made simulations for this quantity after starting with a 
random initial configuration, 
and allowing the  system  to evolve 
according to the rule of NEKIM as described above. Averaging 
has been  taken over 
random initial configurations 
 in a chain of length $L=1000$. The result
is shown on Fig.2. The best
fit has been obtained with $\lambda=1.49(3)$, using $Z=1.75(1)$.
 For comparison, numerical results for the  corresponding
quantity in the Glauber-Ising limiting case are also displayed
on Fig.2. It is worth mentioning, that data for $t\leq 10$ had to
be discarded in both cases; 
power law behaviour is seen only for later times and
this fact does not change if the number of averages taken is increased
even by an order of magnitude. 
\begin{figure}[ht]
  \centerline{\epsfxsize=8cm
                   \epsfbox{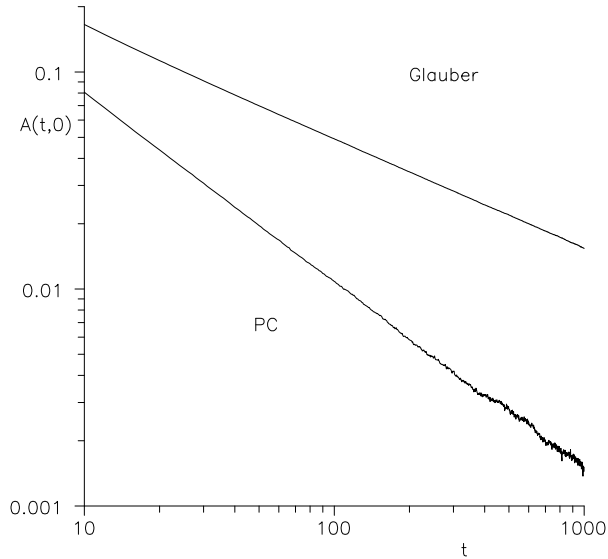}
                    \vspace*{4mm}
             }
   \caption{Time dependence of the local autocorrelation function $A(t,0)$.
$L=1000$ and the number of averages over independent random initial
states was $1.5\times 10^5$ also for the Glauber-Ising case which is
shown for comparison.}
\end{figure}

Following \cite{MBCS96} , we will study now the two-time 
autocorrelation
function for the {\it global} order parameter:
$A^{global}(t_1,t_2)=L<M_{k=0}(t_1)M_{k=0}(t_2)>$
 or rather its normalized form, namely
\be
a(t_1,t_2)=A^{global}(t_1,t_2)/{\sqrt{S(0,t_1)}\sqrt{S(0,t_2)}}
=f(\frac{t_1}{t_2})
\ee
Here $S(0,t)= L<\frac{1}{L}[{\sum_{i}s_i(t)}]^2>$ is the structure
factor at the ferromagnetic peak and the second
 equality follows from scaling assumption (see later in more detail).
Morerover, for $y\rightarrow\infty$,
$f(y)\sim y^{-{(\lambda-d+1-\eta/2)}/Z}$ is the expected power law
behaviour. Nevertheless, if the process is Markovian, the power law
behaviour of 
$f(\frac{t_1}{t_2})$ has to hold for all $t_1>t_2$ as shown in \cite{MBCS96}.

The second moment of the global magnetization (structure factor)
should behave as \cite{bray0}
\begin{equation}
S(0,t)\sim 
t^{{(d-\frac{2\beta}{\nu})}/Z} \sim t^{\frac{2-\eta}{Z}}
\end{equation}
We have found earlier, in \cite{MEOD96} that $\beta=.00(2)$, i.e. even at the
PC point the Ising phase transition is of first order and thus
$S(0,t)\propto t^{1/Z}$. Moreover, via the above applied 
 scaling law $d-2+\eta=\frac{2\beta}{\nu}$,  $\eta=1.0(1)$ follows
at the PC point, too.

Fig.3. shows $a({t_1}/{t_2})$ as a function of ${t_1}/{t_2}$ for 
six different
values of $t_2$, $t_2=3,5,10,32,50,100$.
We have simulated $A^{global}(t_1,t_2)$  while for the
denominator we have used the power law behaviour as indicated above
 with ${1/Z}=.57$. Unfortunately, it is very hard to get 
$A^{global}(t_1,t_2)$  to
a satisfactory accuracy. 
\begin{figure}[ht]
  \centerline{\epsfxsize=8cm
                   \epsfbox{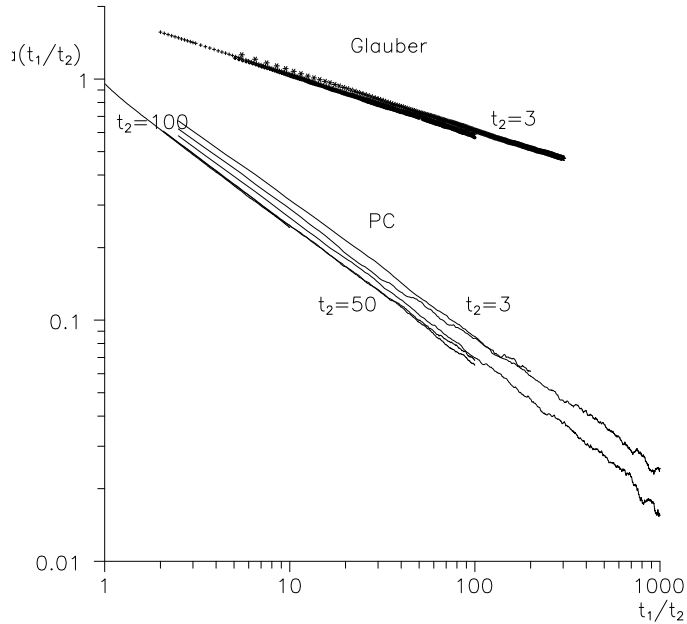}
             }
   \caption{The normalized autocorrelation function $a(t_1/{t_2})$ at the
PC point for six different values of $t_2$ increasing in the downward
direction in the range 3-100. The Galuber case is exhibited again for
comparison. $L=1000$ and the number of averages over independent initial 
states:
$1-2\times10^5$.}
\end{figure}

The reason is probably the fact that in the scaling form of
$A^{global}(t_1,t_2)$ the leading order term is $\propto{({\frac{t_1}{t_2}})}
^{\frac{d-\lambda}{Z}}$ which is non-singular in the present model ( it
is marginal with $d=\lambda$  for the 1d Glauber model while it is singular
for the 2d Ising case).This means that the $k=0$ mode is not special for
the two-time structure factor
 (while it is for the equal-time structure factor) and thus a 
power law behaviour is correction to scaling \cite{mahu}.
In more detail:
the scaling form for the two-time structure factor can be written as
\begin{equation}
<M_{-k}(t_1)M_{k}(t_2)>={t_2}^{\frac{2-\eta}{Z}}{\left[\frac{L(t_1)}{L(t_2)}
\right]}^
{d-\lambda} f_1{(kL(t_1))}+{t_2}^{\frac{2-\eta}{Z}}{\left[\frac{L(t_1)}{L(t_2)}
\right]}^
{\lambda_1} f_2{(kL(t_1))}, \, \, \, t_1,t_2\gg 1, \,\,\, d>\lambda
\end{equation}
where $L(t)\sim t^{1/Z}$ and for $k\rightarrow 0$  $f_1{(kL(t))}
\rightarrow const$.
Moreover since the second term is correction to scaling, 
$\lambda_1 <d-\lambda$. 
 For $t_1=t_2$ we get the usual structure factor and for
$k\rightarrow 0$ without the correction term  this is the form cited in 
\cite{MBCS96} below their eq.(16). In the present case, 
knowing that the singular term
is missing we can say $\lim_{k \rightarrow 0}f_1{(kL(t))}=0$, 
scaling can still be present and $\lambda_1$ 
plays the
role of $d-\lambda$ (and we will use $d-\lambda$ for $\lambda_1$ in the
following even for $\lambda>d$).

Now turning back to our Fig.3, apart from the first three decades in time,
fluctuations hinder drawing any consequence  concerning the
(correction to scaling ) behaviour of $A^{global}(t_1,t_2)$ 
even for averages of order $10^5$. For the quantity $t_1/{t_2}$ this
fact narrows down the interval of analyzable data even more.
Nevertheless, it is clearly seen that the dynamic scaling assumption
expressed in eq.(6) ( and which eventually can be expected to
hold only for $t_1\gg 1,t_2\gg 1$!)
starts to be fulfilled to an accuracy below
$1\%$ only for values  $t_1>t_2 \gtrsim 50$. 
Actually this is not typical:
it has been proposed \cite{Bray} 
 that in case of systems quenched to  their 
critical temperature (and here $T_c=0$) 
universality and scaling may appear in a quite early stage of time
evolution, far from equilibrium, where $\xi(t)\sim t^{1/Z}$ is still small. 
Based
on the scaling relation for such early time intervals, a new way for
measuring static and dynamic exponents has been proposed \cite{bray0,Li}
and applied also for the local autocorrelation function\cite{zheng96}.
Some of our earlier results also show that, indeed,  power law 
behaviour sets in 
for quite early times already. Thus e.g. in 
\cite{MEOD96} concerning  the structure factor 
 $S(0,t)\sim t^x$,  
 the power law behaviour was apparent already for very early times 
and for such low values
of $L$ as $L=128$,  provided the number of averages in the simulation
was high enough (above $10^5$). The 
obtained result, $x=.57=1/Z$ was made use of above.

For the sake of comparison
we have carried out similar simulations of $A^{global}(t_1,t_2)$
for the exactly soluble
Ising-Glauber case ($p_{ex}=0, \delta=0$), some of these are also exhibited
on Fig.3. Here dynamic scaling is fulfilled (to similar accuracy as above)
already for $t_1>t_2 \gtrsim 5$ and the expected power law behaviour
is seen within error.
It is worth mentioning that similar value for $\lambda$, i.e.
$\lambda=1.0$ results from simulations in the whole absorbing region
(thus e.g. for $p_{ex}=.35, \delta=0$).
\begin{figure}[ht]
  \centerline{\epsfxsize 8cm
                   \epsfbox{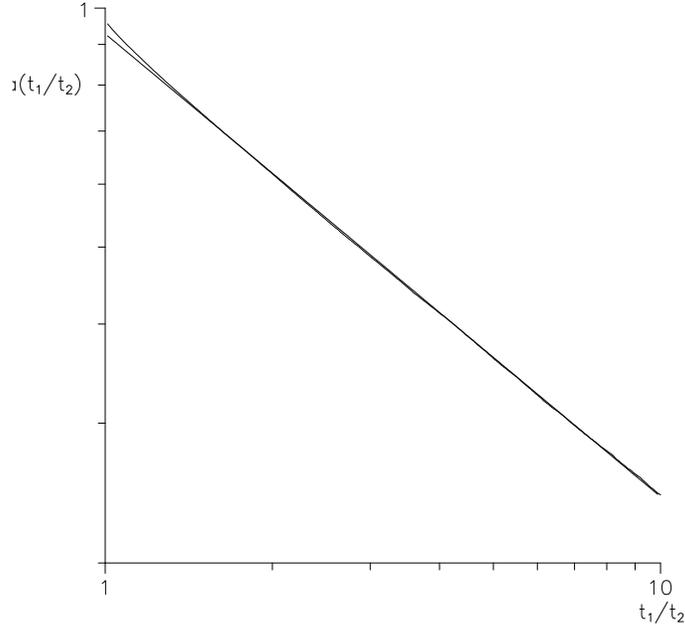}
                   \vspace*{4mm}
             }
   \caption{
The dependence of the normalized autocorrelation function
on ${t_1}/{t_2}$ in its first decade for $t_2=100$. $L=1000$ and the number
of averages was $10^6$. The straight line is best power law fit with
${(\lambda-\frac{\eta}{2})}/Z=.581$}
\end{figure}

In order to establish whether the process is Markovian or not at the PC
point it will be sufficient  to examine
the first decade in the variable $\frac{t_1}{t_2}$ in a region where dynamic
scaling holds. Fig.4 shows the result for the case $t_2=100$ and for
averages over $10^6$ independent initial states, again taking $L=1000$.
For the exponent $\frac{(\lambda-{\eta/2})}{Z}$ the value $.58$ results
as best fit, which, according to eq.(1) should equal $\Theta$. From
here one arrives at $\lambda=1.51(1)$ which is in accord, within error,
with the value obtained above from the local autocorrelation function.
Thus, supposing the Markovian property to hold has led to contradiction
because the measured value of $\Theta$ is $.67(1)$.

The results together with critical exponents obtained earlier in
\cite{MEOD96} are summarized in Table I.
\bigskip
\begin{table}[h]
\begin{tabular}{l|l|l|l|l|l|l|}
     & $\beta$& $\gamma$ & $\nu$ & Z & $\Theta\,\, $ & $\lambda\, $  \\
\hline
Glauber-Ising& 0 & $1/2$ & $1/2$ & $2$   & $1/4$ & $1$ \\
\hline
PC &$ .00(1)$ &.444(2) &$.444(2)$ & 1.75(1) & $.67(1)$ & 1.50(2)\\
\end{tabular}
\caption{\em Simulation data for static and dynamic critical exponents
for NEKIM}
\end{table}

In summary, we have carried out numerical simulations to investigate
the non-equilibrium dynamic critical exponents $\Theta$ and $\lambda$
with the aim to check the Markovian nature of the nonequilibrium
Ising system in 1d at the parity conserving phase transition point
of the phase diagram of NEKIM.
On the basis of of the present results we have been led
to the conclusion that the effect of fluctuations felt
by the spin system at the PC transition is such that
the dynamical process becomes non-Markovian. The difference is
quite pronounced, definitely beyond numerical errors.

\bigskip
{\bf ACKNOWLEDGEMENTS}\\
 The  authors thank C.Sire for useful correspondence.
Support from 
  the Hungarian research fund OTKA ( Nos.
T017493, 027391 and 023552) is gratefully acknowledged.
 The simulations were partially carried out on
the Fujitsu AP1000 parallel supercomputer.

\end{document}